# Wavelet Analysis of Dynamic Behaviors of the Large Interconnected Power System

Samir Avdakovic, Amir Nuhanovic, Mirza Kusljugic, Elvisa Becirovic

**Abstract**— In this paper, the simulation of the disturbance propagation through a large power system is performed on the WSCC 127 bus test system. The signal frequency analysis from several parts of the power system is performed by applying the Wavelet Transform (WT). The results show that this approach provides the system operators with some useful information regarding the identification of the power system low-frequency electromechanical oscillations, the identification of the coherent groups of generators and the insight into the speed retardation of some parts of the power system. The ability to localize the disturbance is based on the disturbance propagation through the power system and the time-frequency analysis performed by using the WT is presented along with detailed physical interpretation of the used approach.

—————— ◆ ——————

## 1 INTRODUCTION

THE electric power system is a complex dynamic system that is constantly exposed to disturbances of different intensity [1]. Despite the best efforts of system planners to provide safe and reliable supply of electricity to customers, several partial or complete power system blackouts have occurred worldwide during the past decade. In order to provide more secure and more reliable power system operation, Wide Area Monitoring Protection and Control Systems (WAMPC) are developed intensively. These systems are based on real time voltage and current measurement in points of special interest in power system (Phasor Measurement Unit devices). This platform enables the real dynamic snapshot of system, higher measurement precision, fast data exchange and alarms during side effects [2], [3], [4], [5], [6]. Electromechanical transients are slower (system) phenomena and are the consequences of physical behaviour of synchronous generators when connected to the network. This system contains different energy storage systems such as rotational machine parts which produce oscillations as a response to the smallest imbalance. In a large power system, one can notice the propagation delay between two system points, actually two machines [7], [8]. The '*Virginia Tech's Frequency Monitoring Network* (FNET)' is developed on the basis of such a phenomenon [9], [10]. The system oscillations which arise from disturbances might cause the unwanted operation of the under-frequency system protection, thus additionally complicating the current status of the power system.

The Wavelet Transform (WT) has been drawing a lot of attention in terms of the signal analysis in the last few years. The main advantage of the WT when compared to other standard techniques is the time-frequency analysis. The detailed analysis of a vast number of previously works on the WT application in the signal analysis of electromechanical oscillation in the power system mainly indicates the analysis and identification of low-frequency electromechanical oscillations, the identification and localizations of disturbances and the identification and the estimation of the active power imbalance [11], [12], [13], [14], [15], [16], [17], [18], [19], [20], [21], [22]. The WT enables direct estimation of the rate of change of weighted average frequency (frequency of the centre of inertia) and disturbance localization thus representing an excellent basis for the improvement of the existing WAMPC systems [20], [21]. The focus of this paper is to analyze the signals following a disturbance at multiple points of a large power system using the WT and represents the continuation of our previous work presented in [20] and [21]. Contrary to the small power system, in a large interconnected power system the time delay in disturbance propagation is evident and it is common for large systems to have more coherent groups of generators. The simulations performed on the WSCC 127 bus test system show that regional platform for synchronized measurements and WT signal analysis provide power system operators with useful information and aids the understanding of the dynamic behaviour of large interconnected power systems after disturbance. Also, the possibility of the disturbance localization, in accordance with the nature of disturbance propagation through the power system and the time-frequency analysis enabled by the WT is presented in the paper along with detailed physical interpretation of the used approach.

This paper is organized as follows: Section 2 briefly describes the phenomenon of the disturbance propagation and dynamic behavior of large interconnected power systems. The wavelet multi-scale analysis of the signals of low-frequency electromechanical oscillations is described in Section 3. The results of the simulation and analysis performed by using the WSCC 127 bus test system are presented in Section 4. The conclusions based on the performed analysis are given in the final Section of this paper.

## 2 ELECTROMECHANICAL WAVE PROPAGATION AND DYNAMIC BEHAVIOR OF LARGE POWER SYSTEM

In order to understand the dynamics and stability of the power system better, it is necessary to know the concept of the dynamic events in the power system. The dynamic characteristic of the power system represents the response of the system to certain disturbances such as: the loss of the generation unit, lightning, short circuits, etc. The important classification of the power system dynamic characteristics is based on the time and frequency range of certain events. The system response to an overvoltage occurrence is the fastest in terms of the time range and it is the result of a changeable phenomenon caused by the atmospheric discharges or the manipulation of high-voltage transmission lines (from microseconds to milliseconds). The electromagnetically dynamic response to the disturbances in machine windings (milliseconds) is a bit slower. The electromechanical system response which results in the oscillations of the rotating machine parts after the system disturbances (up to several seconds) is much slower. The thermodynamic response is the slowest and represents the control of the steam turbine boilers as the result of the frequency variations (minutes or hours). The frequency range of the power system transients is between $10^{-4}$ Hz and up to several kHz. The electromagnetic transients are usually the result of the network configuration changes due to the disconnection or electronic device operation, transient disturbances, etc [1]. The electromechanical transients are slower in time, occur at the system level and are the result of the natural characteristic of the synchronous machines to respond by producing oscillations, with the frequency range up to 5 Hz, when subjected to even the minimal imbalance. The electromechanical wave propagation phenomenon which occurs due to disturbances is described in several books and papers. It is mathematically defined [7], [8] that the change of the frequency or angle propagates in the power system in the wave form. In physics, the wave is defined as '*a disturbance caused by the movement of energy form some source through some medium*' [23], [24]. Any disturbance in the power system, i.e. the loss of the generating units or the loss of the load, will represent the source of an electromechanical wave while its propagation medium will be the electric power network and will have an impact on the angular velocity of the synchronous generators thus the disturbance will propagate through the power system in form of an electromechanical wave.

In real power systems, the frequency instability will occur when the power system is not able to maintain the active power balance in the system that can lead to the frequency collapse. When the system is subjected to the high-intensity disturbances or the series of the cascade element tripping, the frequency instability may occur thus leading to the partial separation of the power system or eventually to the total system blackout. If the active power imbalance in the system persists, usually caused by the loss of the load or generation, the remaining generation units will start to increase or decrease its speed due to the nature of the disturbance. Since the inertia of the generators is different, so that the distribution of its impact results in a disturbance, the generators start to rotate in different speed and generate the voltages of different frequencies. When the transient process is finished, one can expect the unique system frequency – frequency of center of inertia. Large interconnected systems are spread on geographically wide areas and have additionally increased the reliability and security of supply and also enable the exchange and trade of electricity among the interconnected power systems. The interconnection of the power systems additionally complicates the dynamic characteristics of the system. Unlike some connected (national) power systems, which have small transmission impedances due to the short transmission lines and meshed transmission network), large interconnected systems have relatively high impedances of transmission lines between their mutually interconnected systems. This is mainly due to a small number of interconnecting lines with the large length and high impedances. The stronger synchronizing powers exist among the generators in the "weak" interconnections than among the generators in the neighbouring systems, and, as a result, so called "groups of the coherent generators" are formed in these interconnections. Among these coherent groups of generators, the synchronizing powers are weaker and the oscillations of the relative angles among them occur in the dynamic response. These oscillations cause the power oscillations in the interconnection lines [1], [25] and the disturbances can sometimes cause the system oscillations as a whole, with the coherent groups of generators oscillating in the opposite direction of each other, causing certain long-term oscillations of the power among the power system areas. The low-frequency oscillations can decrease the transmission system in such a high manner that any further disturbance might lead to the system collapse. All this emphasizes the importance of the WAMPC systems that have to ensure the quality image of the power system dynamics to the power system operators and to protect the system of any disturbances.

## 3 WAVELET MULTI-SCALE ANALYSIS

Although it represents a relatively new mathematical technique, the WT theory basis can be found in a vast number of books and papers [26], [27], [28], [29], [30]. The wavelet is the wave function with the compact support. It is named the wave because of the oscillatory nature of the wave, and "small" because the final domain is different from zero (of the compact support). The scaling and translation of the basic wavelet (mother) both define the wavelet basis that represents the wave function of a limited duration. By selecting the scaling and translation parameters, the small parts of the complicated forms can be represented with the higher time resolution, while the smooth parts can be represented with the lower time resolution. The WT application in the signal analysis takes both DWT and CWT into consideration. The DWT is the most commonly used WT and represents the recursive filtering process of the data input on the low-pass and high-pass filters. The approximations are the low-frequency components on the large scale and the details are the high-frequency components on the small scale. The transformation of the function with the wavelets can be interpreted as the function "passing" through the bank of filters. The output(s) are the scaling coefficients (approximations) and wavelet coefficients (details). As previously noted in Section 2, the frequency range of the electromechanical transients in the power system is low and goes up to 5 Hz. So, according to the Nyquist' Theorem, sampling the frequency of 10 Hz is sufficient. The frequency bands corresponding to five decomposition levels of the DWT for the wavelet functions and the signal with the sampling frequency of 10 Hz are divided into D1-D5 details and one final A5 approximation as follows: D1 [5.000–2.500 Hz], D2 [2.500–1.250 Hz], D3 [1.250–0.625 Hz], D4 [0.625–0.312 Hz], D5 [0.312–0.156 Hz] and A5 [0.156 – 0.000 Hz].

## 4 STUDIED SYSTEM

Simulations and analysis of the electromechanical wave propagations in this paper are performed by using the WSCC 127 bus test system. In physics, it represents one of the largest interconnected power systems in the world and geographically covers the area from west Canada, over the central and western part of the USA, and all the way to Mexico. The one-line diagram can be found in [31], and the geographical location of the system in [32]. The total active load of this test system is 60.785 MW. The referent points for the frequency signal analysis are selected from every part of the system and they are as follows: bus 9 (HANFORD 500.00 kV), bus 14 (BIG EDDY 500.00 kV), bus 27 (MIDPOINT 500.00 kV), bus 30 (BENLOMND 345.00 kV), bus 40 (PINTO 345.00 kV), bus 43 (INTERMT 345.00 kV), bus 71 (MIRALOMA 500.00 kV), bus 82 (SYLMAR S 230.00 kV), bus 90 (TEVATR 500.00 kV), bus 95 (ROUND MT 200.00 kV), bus 110 (STAF 230.00 kV) and bus 118 (RINALDI 230.00 kV). The sampling frequency is 10 Hz. After the simulation of the loss of the generation on HAYNES3G bus 113 in 1.0 sec, with the active power loss of 325 MW, the frequency deviations on selected buses are represented in Fig. 1.

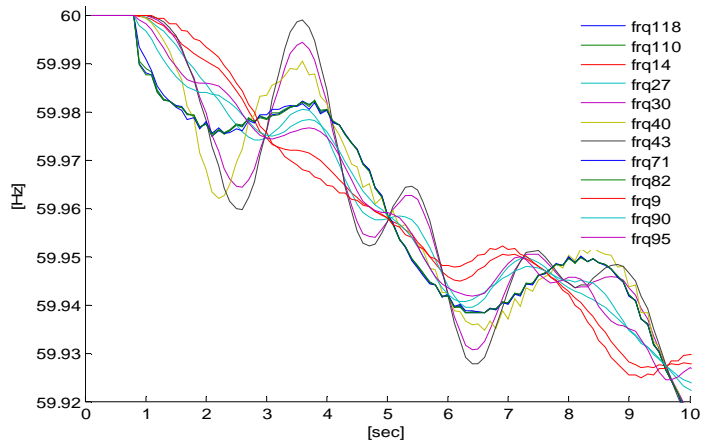

Fig. 1. Frequency deviation/propagation after a loss of generation units HAYNES3G at node 113 of the WSCC 127 bus test system.

As it can be seen in Fig. 1, the frequency deviation after the loss of the generation will first be evident on the buses that are electrically closer to the disturbance location, in particular buses 118, 110, 82 and 71. These buses are geographically located in the southern part of the system, so are the HAYNES3G generation units at node 113, so they are electrically closest to the disturbance location. After some time delay, the frequency starts to decrease on the other buses, first in the central and middle part of the power system and then in buses 14 and 9, which are geographically located in the northern part of the power system. Also it is possible to note that frequencies in buses 118, 110, 82 and 71 oscillate very similarly, so it might indicate the existence of the coherent groups of generators in that part of the power system [1]. It is also possible to notice the same as far as buses 90, 95 and 27, and also for buses 43, 40 and 30, and also for buses 14 and 9 in the northern part of the power system.

The results of the signal analysis from Fig. 1, where the DWT, the Db4 wavelet function with five decomposition levels, is used are presented on Fig. 2. The Db4 transform has four wavelet and scaling function coefficients [21]. The components of the frequency signal from Fig.1 are given in Fig. 2 for the frequency range [5.000–2.500 Hz], which represents the first level of the decomposition of the analyzed signals (D1).

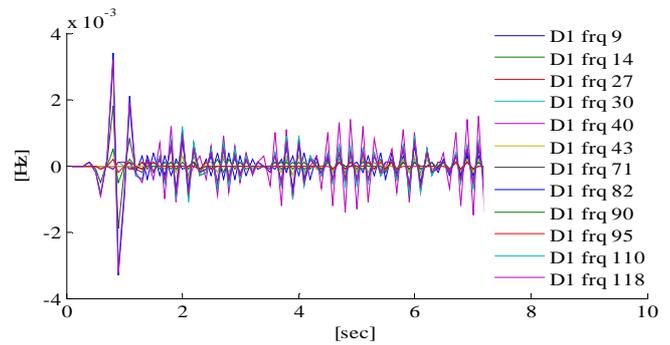

Fig 2. Components of frequency signal from Fig. 1 for frequency range [5.000 – 2.500 Hz] – D1.

The first level of the signal decomposition with the DWT is usually used for the identification of the event start or the identification of the signal disturbance. It is obvious from Fig. 2 that the simulated loss of the generation in this frequency range can highly influence the system points that are closest to the disturbance location (node 82 and 118). This is manifested as the "small" frequency sag and the DWT identifies this as the minor transient. Also, along with different D1 amplitudes of the components of the analyzed signals, a certain time delay is evident due to the electromechanical wave propagation through the system. When zooming Fig. 2 in 1.4 sec of the simulation time for a better view and observing the first decomposition level signals in the form of energies (D12), one can see that the energy values of the analyzed signals in the first decomposition level are the largest in vicinity buses - buses no. 82, 118, 110 and 71 (Fig. 3).

The degradation of the other signals will occur later as the electromechanical wave propagates from one point to another. It is the time-frequency analysis, provided by the WT and the nature of the electromechanical propagation delay, that enables the disturbance localization in the power system as it is shown in Fig. 3. The D1-D5 frequency ranges of the signal decomposition levels are the ranges of the low-frequency electromechanical oscillations and in these signals all dominant oscillatory modes will be identified. The inter-area oscillations require a special operators' attention in the large interconnected power systems. The DWT signals obtained in the 5th level of the signal decomposition are shown in Fig. 4 with the frequency range as defined previously. It is obvious that after some time, and due to the disturbance propagation through the power system, certain (D5) components of the analyzed frequency signals will oscillate contrary to the other components. This implies the existence of the coherent groups of generators that will oscillate in the opposite. The simulation results show that the D5 components of the frequency signals from buses 30, 40, 43, 71, 82, 110 and 118 (the buses in the central and southern part) oscillate in the opposition to the D5 components of the frequency signals from buses 9 and 14 (the buses in the northern part of the system). It is possible to see that the D5 frequency components from buses 27, 90 and 95, which are the central buses, after certain time oscillate with the lower amplitude than the other observed buses.

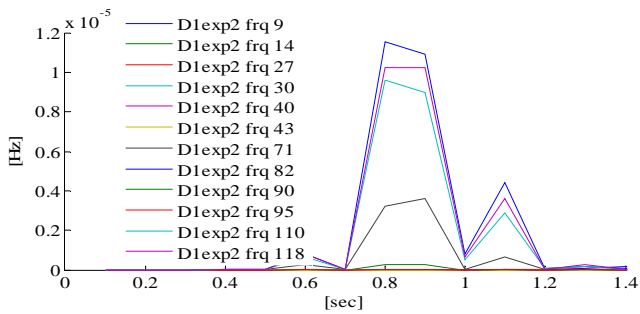

Fig. 3. Energy values D1 of components of frequency signals from Fig. 4 in time period up to 1.4 sec.

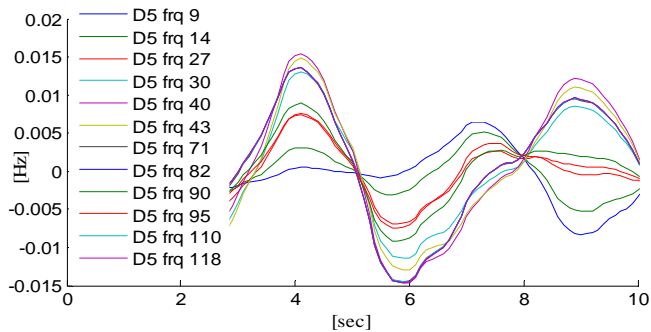

Fig. 4. Components of frequency signals from Fig. 1 for frequency range [0.312 – 0.156 Hz] – D5.

Proper identification of such power system behaviour is of the highest importance to the power system operators. Further, the low-frequency components of the analyzed signals for the frequency range of [0.156–0.000 Hz] are shown in Fig. 5. They represent a good estimation of weigthed average center of inertia and are applicable for total active power imbalance of the power system [21]. But it is obvious that the low-frequency components of the analyzed signals, selected from different parts of the power system, have different propagation paths (Fig. 5). This also points on the existence of coherent groups of generators in power system and, as one can see from Fig. 5, different part of system during transient process retard in somewhat different speeds (dA5/dt).

## 4 CONCLUSIONS

After the disturbance occurrence in large interconnected power systems the electromechanical wave propagation will occur within a certain time delay, which is mainly determined by the power system configuration and the inertia of the system as a whole. As it propagates through the power system, the disturbance causes oscillations, and after a certain period of time, the inter-area oscillations among the coherent groups of generators will arise in the system or in case of active power imbalance the nominal value of system frequency will be deteriorated. Whereas several coherent groups of generators exist in the large interconnected power system, as well as in the analyzed WSCC 127 bus test system, it is shown that, after a certain period of time, the low-frequency oscillations among those coherent groups may occur and complicate the status of the system additionally. In this paper, by using the WT and physicallity of dynamic behaviour of large interconnected power systems, the analysis of frequency signals that might be available from WAMS is performed and simulated on test system.

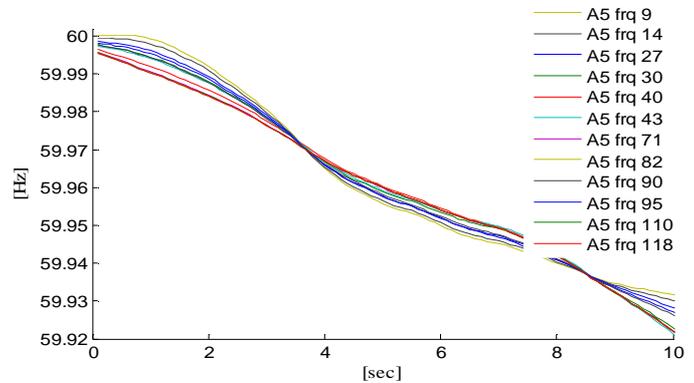

Fig. 5. Components of frequency signal from Fig. 1 for frequency range [0.156 – 0.000 Hz] – A5.

The results of analysis shown the possibility of the disturbance localization by applying the analysis of the energy values of the signals in the first decomposition level (frequency range 2.5–5 Hz), based on the nature of the electromechanical wave propagation and the WT time-frequency analysis. Also, it is shown that this approach enables identification of inter-area oscillations and coherent groups of generators. Additionally, this approach enables an estimation of rate of change of frequency in different parts of interconnected power systems. The platform of synchronized measurement and signal processing by using WT in large interconnected power systems will provide multipurpose information to system operators unlike currently available monitoring systems.

## Authors Biography

Samir Avdaković received B.Eng. and M.Sc. from the Faculty of Electrical Engineering, University of Tuzla and currently he is a PhD candidate. He is working in the Department for Strategic Development in EPC Elektroprivreda B&H. His research interests are: power system analysis, power system dynamics and stability, WAMPCS and signal processing.

Amir Nuhanović is currently a associate Professor in the Department of Power Systems Analysis at the Faculty of Electrical Engineering, University of Tuzla. His research interests are: power system optimization and analysis, power system operation and stability, operations research and numerical techniques.

Mirza Kušljugić is currently a Professor in the Department of Power Systems Analysis at the Faculty of Electrical Engineering, University of Tuzla. His research interests are: power system analysis, power system dynamics and stability, energy efficiency

Elvisa Bećirović received B.Eng. and M.Sc. from the Faculty of Electrical Engineering, University of Sarajevo and University of Tuzla, respectively. Currently she is a PhD candidate at the Faculty of Electrical Engineering, University of Zagreb. She is working in the Department for Strategic Development in EPC Elektroprivreda B&H.